\pdfoutput=1
\documentclass[10pt, conference, final, hyphens]{IEEEtran}

\PassOptionsToPackage{hyphens}{url}
\usepackage{cite}
\usepackage[cmex10]{amsmath}
\usepackage{algpseudocode}

\usepackage{balance}
\usepackage{multirow}
\usepackage[dvipsnames]{xcolor}

\usepackage{pifont}
\usepackage{tikz}

\newcommand{\ignore}[1]{}

\usepackage[printwatermark]{xwatermark}
\newwatermark[firstpage,color=black!100,angle=0,scale=1,xpos=0,ypos=129]{\small\textnormal{© 2020 IEEE. Personal use of this material is permitted.  Permission from IEEE must be obtained for all other uses, in any current or \\ future media, including reprinting/republishing this material for advertising or promotional purposes, creating new collective works,  \\ for resale or redistribution to servers or lists, or reuse of any copyrighted component  of this work in other works.}}
\newwatermark[allpages,color=black!100,angle=0,scale=1,xpos=0,ypos=-126]{\small 2020 50th Annual IEEE/IFIP International Conference on Dependable Systems and Networks - Supplemental Volume (DSN-S) \\ Accepted version © 2020 IEEE. \url{https://doi.org/10.1109/DSN-S50200.2020.00027}}

\begin{document}
\title{Pitfalls of Provably Secure Systems in Internet\\The Case of Chronos-NTP}

\author{\IEEEauthorblockN{Philipp Jeitner\IEEEauthorrefmark{1}, Haya Shulman\IEEEauthorrefmark{1}, Michael Waidner\IEEEauthorrefmark{1}\IEEEauthorrefmark{2}}
\IEEEauthorblockA{\IEEEauthorrefmark{1}Technical University of Darmstadt, \IEEEauthorrefmark{2}Fraunhofer Institute for Secure Information Technology SIT}
}

\maketitle

\begin{abstract}
The critical role that Network Time Protocol (NTP) plays in the Internet led to multiple efforts to secure it against time-shifting attacks. A recent proposal for enhancing the security of NTP with Chronos against on-path attackers seems the most promising one and is on a standardisation track of the IETF. In this work we demonstrate off-path attacks against Chronos enhanced NTP clients. The weak link is a central security feature of Chronos: The server pool generation mechanism using DNS. We show that the insecurity of DNS allows to subvert the security of Chronos making the time-shifting attacks against Chronos-NTP even easier than attacks against plain NTP.
\end{abstract}

\section{Introduction}

Network Time Protocol (NTP) is one of the core Internet protocols, used to synchronise time on Internet systems. 
Recently \cite{malhotra_attacking_2016} demonstrated a proof of concept time shifting attacks against NTP.
The idea was to exploit overlapping IPv4 fragments, whereby a shifted time provided in the attacker's fragment would overwrite the time provided by the real NTP server in a response sent to the client. However, this attack is limited because it depends on various factors not typically present in the Internet, mainly minimum MTU support down to 68 bytes and an IP defragmentation algorithm not used by any current operating system.

Due to the critical role that NTP plays in the Internet there was an immediate followup work, which devised enhancements to NTP, called Chronos \cite{ntp:chronos}, to block the attack. Chronos leverages ideas from distributed computing on clock synchronisation in the presence of Byzantine adversaries and is designed to provide security even against strong Man-in-the-Middle (MitM) attackers and corrupted NTP servers. 

\subsubsection*{Our contribution}
In this work, we present an DNS-poisoning based attack on NTP, which is not only more practical than attacking NTP directly, but works even against the Chronos enhanced NTP clients. This attack is part of research on security of NTP with DNS \cite{ntp-over-dns}.

\section{DNS Cache-Poisoning Attacks}
DNS cache poisoning \cite{cns:frag:dns,shulman2015towards,brandt2018domain} allows off-path attackers to gain MitM capabilities for some target victim domain. This allows the attacker to intercept any communication with the victim domain, such as web, email, FTP or in this case -- NTP. Off-path cache poisoning attacks were demonstrated using BGP hijacking and defragmentation poisoning. 

BGP hijacking places the attacker in a MitM position for the victim network\cite{dns:venezuela,myetherwallet}.

\subsection{Attacking DNS with De-fragmentation Poisoning}
IP Defragmentation cache-poisoning allows off-path attackers to hijack the DNS without the necessity to control BGP routers for prefix hijacking. In 2011 \cite{cns:frag:dns} demonstrated practical DNS cache poisoning attacks that use fragmented IPv4 packets. The idea is to inject a spoofed fragment into the IP defragmentation cache that is to be reassembled with the real fragment from the nameserver. 
 
In \cite{ntp-over-dns} we evaluated fragmentation support of {\it pool.ntp.org} nameservers as well as DNS resolvers the internet. We found that 16 out of 30 nameservers for {\it pool.ntp.org} fragment DNS responses down to a MTU of 548 bytes while not supporting DNSSEC. There are also challenges with DNSSEC deployment \cite{herzberg2013towards,dai2016dnssec,shulman2017one}. Furthermore, using an ad-network based study we found that 90\% of resolvers accept fragments of some size and 64\% even the tiniest possible fragment size of 68 bytes MTU. 

Additionally we found that DNS resolvers are often shared by different systems in the internet. This simplifies fragmentation-based DNS poisoning because the attacker can execute the DNS-poisoning and time-shifting attacks independently, picking a protocol which allows for easy triggering of DNS queries and yields bigger DNS responses to get over the minimum MTU size of the namserver. For 14\% of DNS resolvers used by web clients attackers we were able to trigger queries via either SMTP servers or open resolvers.

\section{Chronos -- Security Enhanced NTP}
Chronos is a formally verified, backwards-compatible NTP client which aims to prevent attacks against NTP using distributed-computing techniques\cite{ntp:chronos}. In contrast to traditional NTP which queries few (typically up to 4) NTP servers, Chronos queries time from multiple NTP servers, applies a secure algorithm for eliminating suspicious responses and averages the time over the valid responses. The authors demonstrate that in order to shift time on a Chronos NTP client by 100ms a strong MitM attacker would need 20 years of effort.

Chronos {\em implicitly} assumes that the attacks on NTP are launched when the NTP client queries the NTP servers in the Internet for time. It's attacker model is that of a MitM attacker that changes some responses to contain shifted time or an attacker that corrupts some of the NTP servers. The security of Chronos is guaranteed if the majority of the queried NTP servers provide correct time. The idea is that all the responses are fed to an algorithm which calculates time by discarding the bottom and top third of responses and sets the time to the average value of the surviving time samples.

Therefore, compared to traditional NTP, Chronos changes NTP clients in two ways. First, by using a bigger pool of upstream NTP servers and second, by replacing the clock select algorithm with one which is formally verified to be secure against a minority of malicious responses.

\section{Attacking Chronos with DNS}
\label{sec:attackingchronos}
In this work we identify the ``achilles heel'' of Chronos: generating a pool of NTP servers which satisfies it's assumption of a honest majority of servers.
The idea is to collect a set of roughly hundred NTP servers, from which NTP servers are selected at random and queried for time samples. 
This pool is generated by querying the {\it pool.ntp.org} domain hourly for 24 hours, yielding a number of $96$ NTP servers when each DNS response contains $4$ NTP servers as in the case of {\it pool.ntp.org}.

Chronos' server pool generation method is flawed in a way which ironically simplifies a DNS-based attack compared to traditional NTP clients. Because the DNS is queried 24 times, the attacker has potentially up to 24 tries to succeed with an cache-poisoning attack instead of just once in case of a traditional NTP client. We show that if the cache-poisoning attack succeeds until or during the 12th DNS request, the attacks still controls more than $\frac{2}{3}$ of the addresses included in the server pool, the definite maximum of servers required for a time-shifting attack on Chronos. How the cache poisoning is done -- via BGP hijacking or fragment injection -- and whether the query is triggered by Chronos directly of via a third party system like SMTP is not important for this attack to work.

\begin{figure}[t!]
    \centering
    \includegraphics[width=0.45\textwidth]{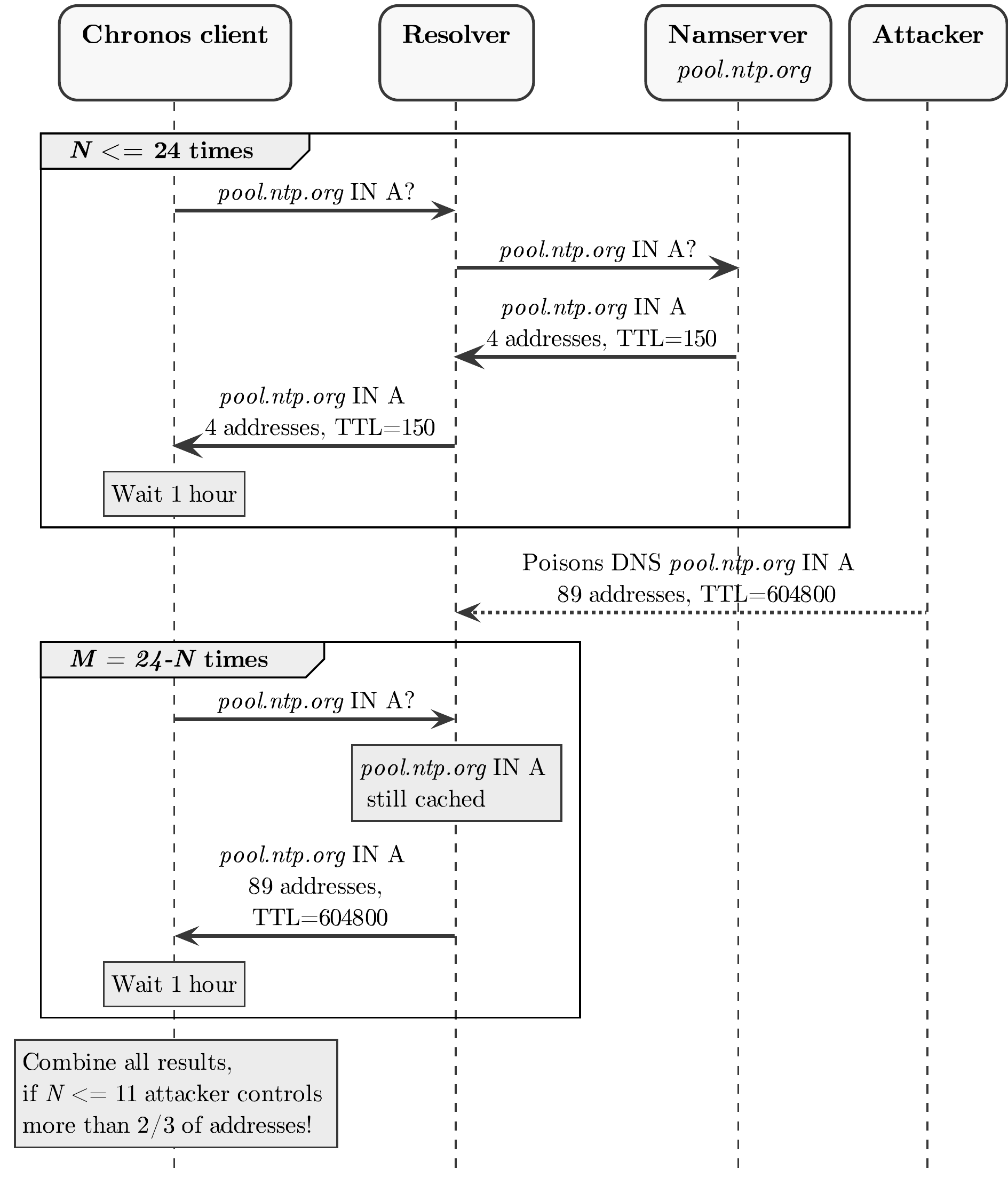}
    \caption{DNS poisoning attack on chronos.}
    \label{fig:chronos_attack}
\vspace{-13pt}
\end{figure}

What simplifies the attack on Chronos is the fact that other than the benign {\it pool.ntp.org} namservers, (1) the attacker can fit more addresses in a single DNS query (up to 89 for a single non-fragmented DNS response) and (2) set the DNS Time-to-live (TTL) to a value bigger than 24 hours, to cause any subsequent queries to be answered from cache, effectively not adding any new server to the pool. When this is done as shown in Figure~\ref{fig:chronos_attack}, the resulting pool will consist of up to $4 \cdot 11 = 44$ benign and $89$ malicious NTP servers which is a $\frac{2}{3}$ majority for the attacker. To succeed with this attack, the attacker therefore only needs to successfully attack the DNS once out of 12 queries during the first 11 hours of Chronos server pool generation mechanism.

\section{Conclusions}
Our attack demonstrates the risks of building security on insecure Internet foundations like DNS, and the risks of analysing security in isolation. The Chronos server pool-generation method can be improved to limit the impact of our attack -- by not allowing more than 4 addresses in a single DNS reply and discarding responses with high TTL values. However, even with these mitigations, the dependency on the insecure DNS still remains and allows way easier attacks than Chronos was originally designed for -- e.g., when the attacker manages to hijack the victim's DNS for a period of 24 hours. To devise secure solutions based on DNS we recommend proposals for generating distributed consensus in a secure way, such as that proposed in \cite{jeitner:consensus:dsn:20}.

\IEEEpeerreviewmaketitle
\bibliographystyle{IEEEtran}
\bibliography{ntp,refs,NetSec}
\balance

\end{document}